\documentclass[intlimits,twoside,a4paper]{article}

\usepackage{amsmath,amssymb}

\usepackage{graphicx}
\usepackage{color}

\usepackage[T2A]{fontenc}
\usepackage[cp1251]{inputenc}

\usepackage{cmpj2}

\issue{2015}{18}{4}{43602}
\doinumber{10.5488/CMP.18.43602}

\title[Composition dependence of the properties of water-methanol mixtures]
{On the composition dependence of thermodynamic, dynamic and dielectric properties of
water-methanol model mixtures. Molecular dynamics \\ simulation results}

\author[E. Galicia-Andr\'{e}s \textsl{et al.}]{E. Galicia-Andr\'{e}s\refaddr{label1},
H. Dominguez\refaddr{label2},
L. Pusztai\refaddr{label3}, O. Pizio\refaddr{label1}\thanks{E-mail: oapizio@gmail.com}}
\addresses{
\addr{label1}Instituto de Qu\'{i}mica, Universidad Nacional Aut\'{o}noma de M\'{e}xico,\\
Circuito Exterior, 04510, M\'{e}xico, D.F., M\'{e}xico
\addr{label2}Instituto de Investigaciones en Materiales,
Universidad Nacional Aut\'{o}noma de M\'{e}xico, \\
Circuito Exterior, 04510, M\'{e}xico, D.F., M\'{e}xico
\addr{label3}Wigner Research Centre for Physics, Hungarian Academy of Sciences, \\
Konkoly Thege ut. 29-33, Budapest, H-1121, Hungary}

\date{Received May 28, 2015, in final form July 29, 2015}
\authorcopyright{E. Galicia-Andr\'{e}s, H. Dominguez, L. Pusztai, O. Pizio, 2015}

\begin{document}
\maketitle

\begin{abstract}
We have investigated thermodynamic and dynamic properties as well as the dielectric constant
of water-metha\-nol model mixtures in the entire range of composition by using constant pressure
molecular dynamics simulations at ambient conditions.
The SPC/E and TIP4P/Ew water models are used in combination with the OPLS united atom
modelling for methanol. Changes of the average number of hydrogen
bonds between particles of different species and of the fractions of differently bonded molecules
are put in correspondence with the behavior of excess mixing volume and enthalpy,
of self-diffusion coefficients and rotational relaxation times.
From the detailed analyses of the results obtained in this work,
we conclude that an improvement of the description of an ample set of properties of water-methanol
mixtures can possibly be reached, if a more sophisticated, carefully parameterized, e.g., all atom,
model for methanol is used.
Moreover, exploration of parametrization of the methanol force field, with simultaneous
application of different combination rules for methanol-water cross interactions, is required.


\pacs 61.20.-p, 61.20-Gy, 61.20.Ja, 65.20.Jk

\end{abstract}

\section{Introduction}

In the very recent work from this laboratory we reported a detailed analysis of
molecular dynamics computer simulation data for the microscopic structure of water-methanol mixtures
in the entire composition range, and performed comparisons with experimental results
obtained using X-ray diffraction  \cite{pusztai1}. Our principal focus in that publication was
on the changes, with composition, of the experimental and theoretical total structure factors
of mixtures considered at room temperature and ambient pressure. On the molecular
dynamics (MD) side, we analyzed the SPC/E  \cite{spce} and the TIP4P/Ew  \cite{tip4p-ew} water models
combined with the OPLS/AA (all atom) methanol model  \cite{opls-aa}.
The all atom modelling for methanol was used in that study because it is intrinsically required by the
procedure to perform comparisons with the experimental total structure factors.

Encouraged by an overall satisfactory performance of the MD
simulation data in the description of the experimental trends, in the present work our
primary objective is to extend our previous study by exploring the results for a more comprehensive
set of properties of water-methanol mixtures, besides the microscopic structure.
Namely, we focus on thermodynamic, dynamic and dielectric
properties obtained from our MD simulations and perform comparisons with
the available experimental data.
Results of computer simulations from other authors are included as well, aiming at a
more comprehensive insight into the dependence of properties of mixtures in question
on the conditions of computer modelling.

There have been very many studies of  water-alcohol mixtures, and of water-methanol mixtures
in particular,
with the use of computer simulation methods since the pioneering
works by Jorgensen and Madura  \cite{jorgensen1} and by Okazaki et al. \cite{okazaki1,okazaki2},
principally focused on  ``infinite dilution'' solutions by considering a single  methanol molecule
in a set of water molecules. The results were obtained using Monte Carlo (MC) simulations in the NPT
and NVT ensemble, respectively. To our best knowledge, Ferrario et al.  \cite{ferrario} and
Tanaka with Gubbins  \cite{tanaka} were the first to report thermodynamic and dynamic properties
of these systems in the entire composition range from MD and MC
simulations, respectively. Since then, much knowledge has been accumulated about the properties
of water-methanol mixtures, see e.g.,  \cite{laaksonen,gabor1,gabor2,gabor3}.
In particular, one of the more recent works on the topic,
with the principal focus on dynamic properties of water-alcohol mixtures (methanol, ethanol
and 1-propanol) using the TIP4P water model  \cite{tip4p} in combination with the OPLS/AA  model
for alcohols  \cite{opls-aa} has been published by Wensink et al.  \cite{wensink}.
On the other hand, Guevara-Carrion et al.  \cite{vrabec1} explored the SPC/E and TIP4P-2005 \cite{vega}
water models in combination with a united atom type methanol model  \cite{vrabec2} for mixtures, reporting excess mixing volume
and enthalpy, self-diffusion coefficients, shear viscosity and power spectra.
Recent works on water-methanol mixtures include detailed studies of the mixing volume and mixing enthalpy
  \cite{nezbeda1,nezbeda2}, using the NPT MC simulations.

Analyses of various results emerging from computer simulations by using non-polarizable water models
and different versions of force fields for alcohols in comparison with experimental data
clearly show the necessity of further work. Namely, it seems necessary to reach a systematic, more profound
understanding of strengths and drawbacks of different models and provide a design
aiming at a better agreement with experiments, desirably for an ample set of  properties of mixtures of
interest, rather than for a single particular property.

In this respect, in the present work we would like to report thermodynamic properties of mixing
in terms of excess mixing volume and excess mixing
enthalpy, a set of dynamic properties that include
the self-diffusion coefficients and rotational relaxation times, as well as the dielectric constant,
all together with the detailed analysis of hydrogen bonding within species and
between them, in the entire range of composition,
starting from pure water and terminating by pure methanol. Whenever possible, we discuss the
results of other authors and make comparisons with experimental data from literature.
We restrict our attention to two different water models and the OPLS/UA (united atom) potential model
of methanol in order to establish their performance. The
OPLS/UA model for methanol and other more complex
alcohols is popular in comparison with the more detailed OPLS/AA model due to computational efficiency.

Several recent efforts have focused on the parametrization of different properties of water,
see e.g.,  \cite{alejandre1,alejandre2}, targeting to improve the description of the dielectric constant and
of the density anomaly.  A similar approach has been applied to methanol
at the united atom level, see   \cite{dominguez}, targeting the dielectric constant in particular.
The strategy of the proposed parametrization includes scaling of site charges with the dielectric
constant as a target, and scaling of the energy of non-bonded
interactions leading to a corrected surface tension as subsequent steps. The critical temperature
is used for control as well. The parametrization can be done in cycles to yield correct values
for the targets. On the other hand, Schnabel et al.  \cite{vrabec2} developed a successful parametrization
of the OPLS/UA type model for methanol using vapor-liquid equilibrium data, focusing on the correct
description of the hydrogen bonds statistics to correctly reproduce the nuclear magnetic resonance
spectroscopic data.

\section{Simulation details}

All our simulations of water-methanol mixtures models were performed in the isothermal-isobaric ensemble
at ambient pressure. The present work involves exploration of two commonly used water models,
namely the SPC/E   \cite{spce} and the TIP4P/Ew ones  \cite{tip4p-ew}.
On the other hand, the OPLS/UA (united atom) rigid nonpolarizable model for
methanol  \cite{jorgensen2} in the present work is taken with
the adjusted parameters of~\cite{vanleeuwen}.
The OPLS/UA model for methanol has been designed so that the parameters of the
nonbonded potentials between
different types of atoms satisfy the geometric combination rules  \cite{jorgensen2}.
However, in the work of van Leeuwen and Smith  \cite{vanleeuwen},
new parameters for nonbonded potentials were designed and standard Lorentz-Berthelot
(LB) combination rules,
i.e., $\epsilon_{ij} =(\epsilon_{ii} \epsilon_{jj})^{1/2}$ and $\sigma_{ij}=(\sigma_{ii}+\sigma_{jj})/2$,
were used to calculate the liquid-vapor coexistence curve of methanol and a bit later
of other alkanols  \cite{monica2}. Again, the LB combination rules were used for methanol
in  \cite{veldhuizen}.

This is a setup similar to the one recently used
by Mou\v{c}ka and Nezbeda  \cite{nezbeda2},
and we just apply the SPC/E and TIP4P/Ew models instead of the TIP4P potential considered by these authors.
Moreover, we have not done any deviation from the LB combination rules, in contrast to \cite{nezbeda2}.
All the simulations of water-OPLS/UA version of the models, unless specified, were performed using the DLPOLY Classic package  \cite{forest}.
We used the Berendsen thermostat and
barostat with $\tau_T = 0.1$~ps and $\tau_P = 2.0$~ps, the running timestep was 0.002~ps.
As common, periodic boundary conditions were used. The nonbonded interactions were cut-off at 11~{\AA},
whereas the long-range interactions were handled by the
Ewald method with a precision of $10^{-5}$. In order to maintain the geometry of water and methanol
molecules, the SHAKE algorithm was used.
For all simulations, the initial configurations were constructed by using from 2000 to 3000 molecules
of the two species placed randomly in a cubic box.
After equilibration, several sets of simulation runs were performed, each lasting for 6--8~ns, with restart
from the previous configuration,  to obtain the averages for data analysis.
Actually, the overall length of simulation was dictated by the stability of internal energy, density and
the dielectric constant,
so that the runs were not shorter than 20~ns.
The precise numbers of molecules of both species used in our simulations are given in table~\ref{tab-simulation-details}.
Changes of the mixture compositions (as well as the compositions themselves) throughout this study are described
by $X_\textrm{m}$ (molar fraction of methanol); its values are given in table~\ref{tab-simulation-details} as well. In addition, we provide
the  weight concentration of methanol for convenience of the reader.

\begin{table}
\caption{\label{tab-simulation-details} The number of molecules of water and methanol in MD simulations,
molar fraction and weight concentration of methanol.}
\vspace{2ex}
\begin{center}
\begin{tabular}{cccc}
\hline\hline
\hspace{5mm} $N_\textrm{w}$ \hspace{5mm} & \hspace{5mm} $N_\textrm{m}$ \hspace{5mm} & \hspace{5mm} \! $X_\textrm{m}$ \hspace{5mm} & \hspace{5mm} wt$_\textrm{m}$ (\%) \hspace{5mm} \\ [0.5ex]
\hline
2000   &0     & 0    & 0   \\
2000   &125   & 0.06 & 10  \\
2000   &281   & 0.12 & 20  \\
1898   &457   & 0.19 & 30  \\
1800   &674   & 0.27 & 40  \\
1600   &899   & 0.41 & 50  \\
1600   &1099  & 0.46 & 55  \\
1600   &1349  & 0.51 & 60  \\
1400   &1462  & 0.57 & 65  \\
1200   &1574  & 0.63 & 70  \\
1200   &1855  & 0.63 & 75  \\
1100   &1800  & 0.69 & 80  \\
800    &2230  & 0.76 & 85  \\
700    &2100  & 0.84 & 90  \\
415    &2670  & 0.91 & 95  \\
0      &2100  & 1.00 & 100 \\
\hline\hline
\end{tabular}
\end{center}
\end{table}

\section{Results and discussion}

First we would like to discuss our observations concerning the properties that do not require any reference
to the species forming the mixture (such as overall density, mixing properties and the dielectric constant)
and next proceed to the properties characterizing individual
species
characteristics such as self-diffusion
coefficients and reorientational relaxation times, and discuss hydrogen
bonding in terms of the average number of bonds per molecule of the particular species, cross bonds and bonding
states of molecules of each species.

\subsection{Density, mixing properties and the dielectric constant}

\begin{figure}[!h]
\begin{center}
\includegraphics[width=0.45\textwidth,clip]{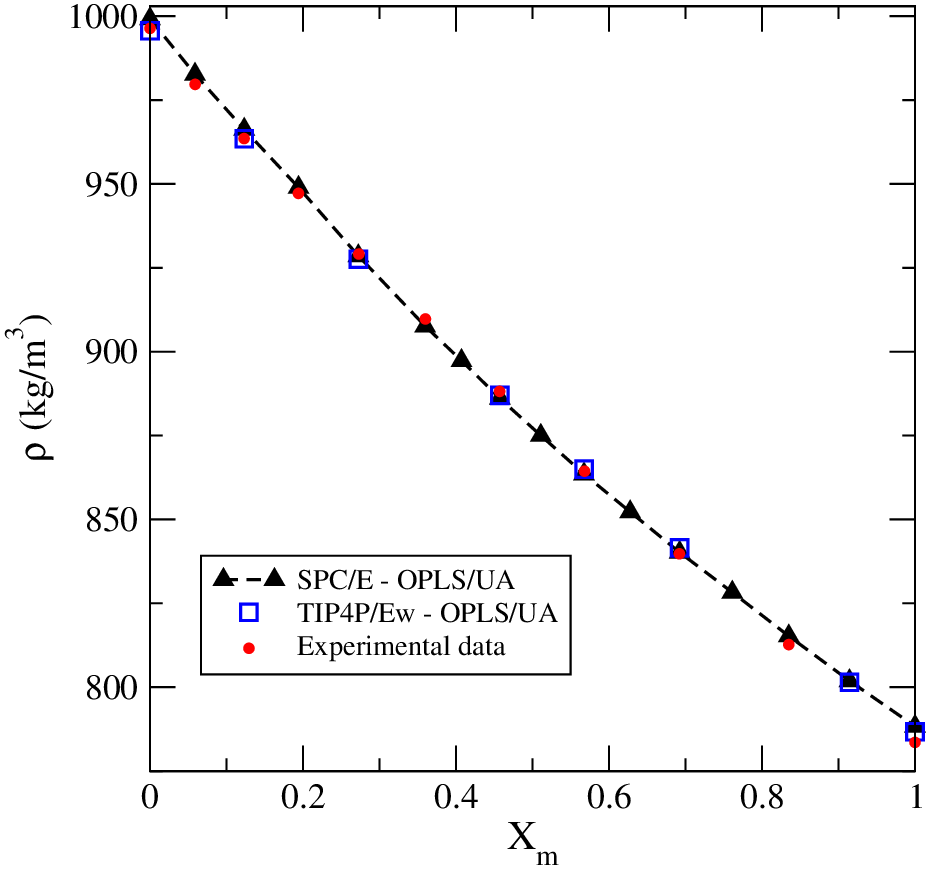}
\end{center}
\vspace{-3mm}
\caption{(Color online) Composition dependence of the water-methanol mixture density from
the present NPT simulations of the SPC/E-OPLS/UA and TIP4P/Ew-OPLS/UA models, together
with experimental results  \cite{tables}. The nomenclature of lines and symbols is given
in the figure.\label{fig1}}
\protect
\end{figure}

Changes of the liquid mixture density with methanol mole fraction for
the two combinations of potential models are shown in figure~\ref{fig1}.
The NPT molecular dynamics simulation data at room temperature (298.15~K)
and at ambient pressure are accompanied by the experimental
values  \cite{tables}. The SPC/E-OPLS/UA model very well reproduces the experimental results. Only
at high methanol fractions, the mixture density is slightly overestimated in comparison with the
experimental data but this disagreement is hardly visible on the scale of the figure.
It is worth mentioning that if the methanol component is described within the OPLS/AA model, then the
density of the mixtures in question is described less satisfactorily: it is underestimated
practically in the entire composition range  \cite{pusztai1}. To conclude with this figure,
it seems that the united atom model for methanol is
reliable to describe changes of the density
on composition of the mixture at room temperature and at ambient pressure.

\begin{figure}[!t]
\begin{center}
\includegraphics[width=0.45\textwidth,clip]{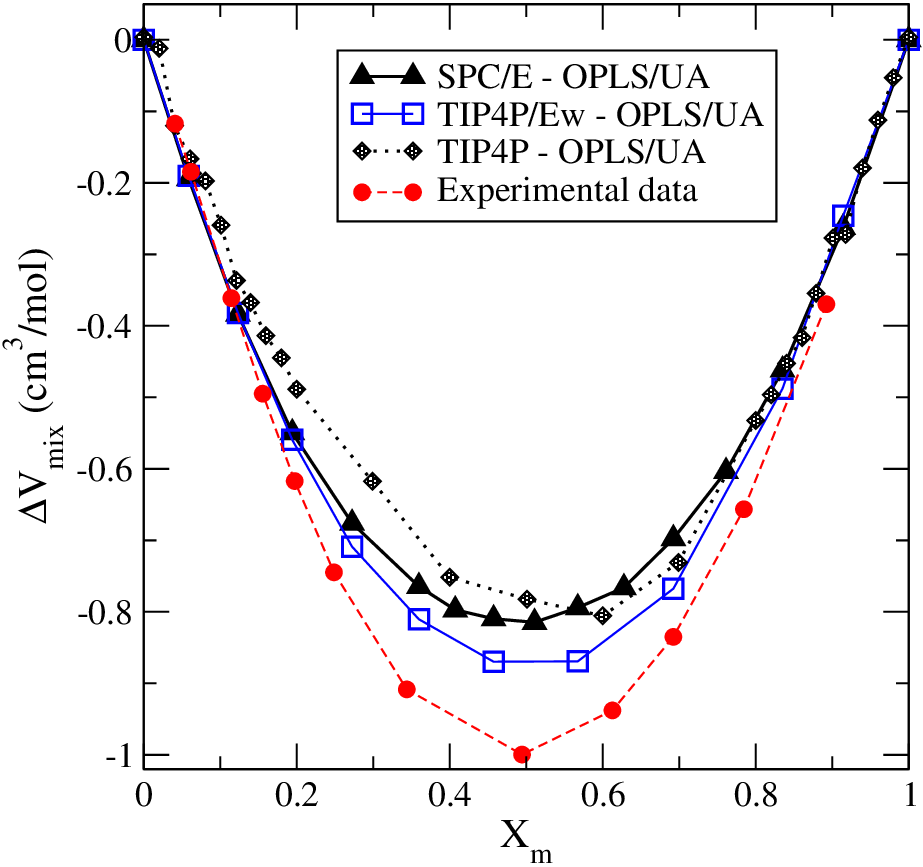}
\end{center}
\vspace{-3mm}
\caption{(Color online) Excess mixing volume of water-methanol mixtures on methanol molar fraction
for united atom modelling from our NPT molecular dynamics and
the results from Monte Carlo simulations  \cite{nezbeda1} for the TIP4P-OPLS/UA model,
versus experimental data  \cite{mcglashan}.\label{fig2}}
\protect
\end{figure}

However, a more sensitive test for the species modelling is illustrated by the calculations of
the excess mixing  volume, $\Delta V_\textrm{mix}$, figure~\ref{fig2}.
It is obtained straightforwardly as: $\Delta V_\textrm{mix} = V_\textrm{mix}-(1-X_\textrm{m})V_\textrm{w}-X_\textrm{m}V_\textrm{m}$.
At the united atom level of modelling, changes of the water model yield a marginal improvement of the
composition dependence of $\Delta V$ with respect to the experimental data \cite{mcglashan}.
Actually, the results are reasonably good  for both water-rich and methanol-rich
mixtures compared to the experiment.
However, at the intermediate region, the applied modelling underestimates the nonideality
of mixtures. Similar trends (to underestimate nonideality) have been observed for the
TIP4P-OPLS/UA
combination of models, if the geometric mixing rules were used for the cross-energies and
cross-diameters  \cite{nezbeda1}, cf. figure~1~(a) of reference  \cite{nezbeda1}.
Moreover, the overall performance of the excess mixing volume coming from simulations
does not improve for different explored sets of deviation parameters from the standard LB
combination rules as shown by  \cite{nezbeda2} for the TIP4P-OPLS/UA model, cf. figure~2 of reference  \cite{nezbeda2}.
Our figure~\ref{fig2} complements the results shown in figure~15 of reference  \cite{vrabec1}. Summarizing our
results and insights coming from the previous simulations in \cite{vrabec1,nezbeda1,nezbeda2}, it
seems that different combinations of water and methanol models yield a satisfactory description of
the excess mixing volume and correctly predict the composition value, $X_\textrm{m} \approx 0.5$ at which
$\Delta V_\textrm{mix}$ is minimum.
However, it is doubtful whether such a moderate success would be preserved for other temperatures and
pressures and/or for other water-alcohol mixtures.
Therefore, accumulation of knowledge about the performance of different models under other conditions
and for other members of water-alcohol mixtures family is of great importance.
On the other hand, it seems necessary to explore parametrization and/or sophistication
of the OPLS/AA modelling,
placing the focus on this particular experimentally measurable property.

\begin{figure}[!b]
\begin{center}
\includegraphics[width=0.45\textwidth,clip]{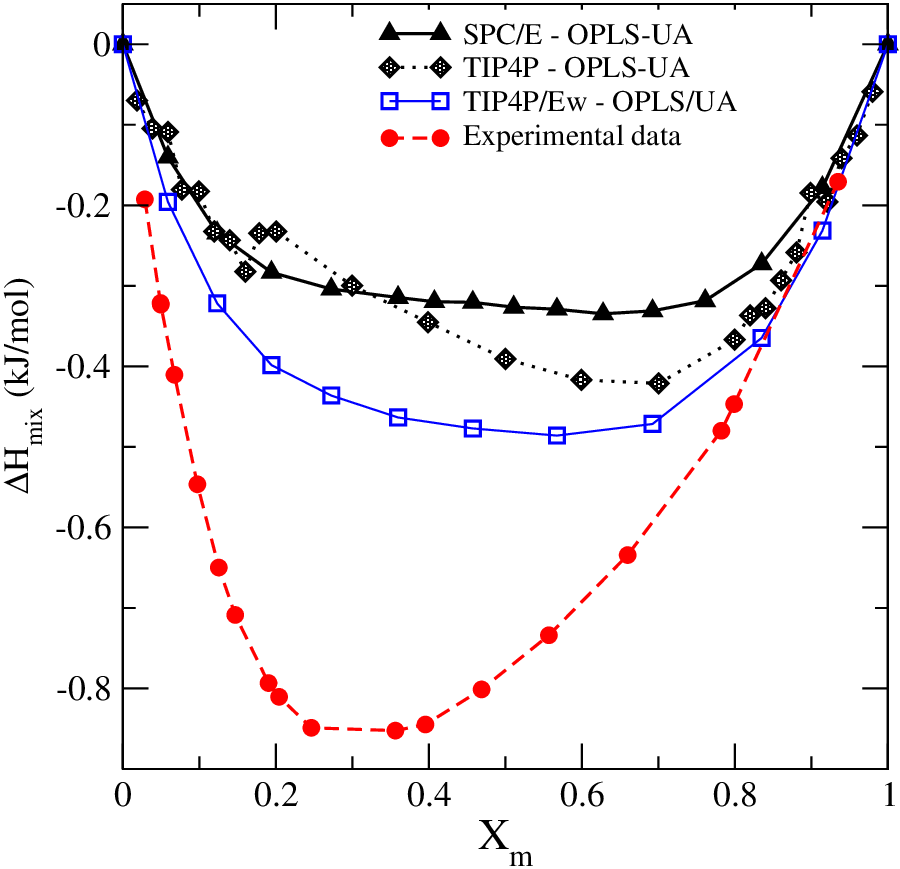}
\end{center}
\vspace{-3mm}
\caption{(Color online) Excess mixing enthalpy of water-methanol mixtures on methanol molar fraction
for united atoms modelling versus experimental data  \cite{lama}. \label{fig3}}
\protect
\end{figure}

The excess enthalpy of mixing, $\Delta H_\textrm{mix}$ is determined as follows:
\begin{equation}
\Delta H_\textrm{mix}= H_\textrm{mix}-(1-X_\textrm{m})H_\textrm{w}-X_\textrm{m}H_\textrm{m}\,,
\end{equation}
where $H_\textrm{mix}$ is $U_\textrm{pot}+PV$.
As we see from the curves in figure~\ref{fig3} and from experimental results  \cite{lama},
the behavior predicted by simulations with
either water models of this study, combined with OPLS/UA, is not satisfactory.
On the methanol-rich side, the behavior
of the theoretical curves seems to be a little better than on the water-rich side. However,
in the entire range of compositions, computer modelling yields much weaker
effects of nonideality compared to the experimental data. Moreover, the minimum of $\Delta H_\textrm{mix}$
from simulations is seen at a higher methanol fraction compared to the experimental results.
These observations are in line
with the previous studies using, for instance, the TIP4P-2005 water model and the parameterized methanol model,
cf. figure~16 of reference  \cite{vrabec1}, and figures~1~(b), 2~(b) of reference~\cite{nezbeda1}. This behavior points out
that the energetic aspects of mixing methanol and water in simulations should be reconsidered.
A growth of the absolute value for
$\Delta H_\textrm{mix}$ can be reached by employing certain deviations to the combination rules  \cite{nezbeda2},
but then the position of the minimum for $\Delta H_\textrm{mix}$ shifts to methanol-rich mixtures, which is
in contradiction to the experimental behavior. Thus, it does not seem that the combination
rules are responsible for this type of failure. One can hope that modelling of methanol
at the all atom level would be profitable to improve an agreement with experimental data for this
particular property. However, another possibility is that nonpolarizable models are intrinsically
incapable of providing a desired agreement for this particular property.
Definitely, other characteristics of the system that involve potential energy,
besides excess mixing enthalpy, should be explored to make stronger conclusions.

\begin{figure}[!b]
\begin{center}
\includegraphics[width=0.45\textwidth,clip]{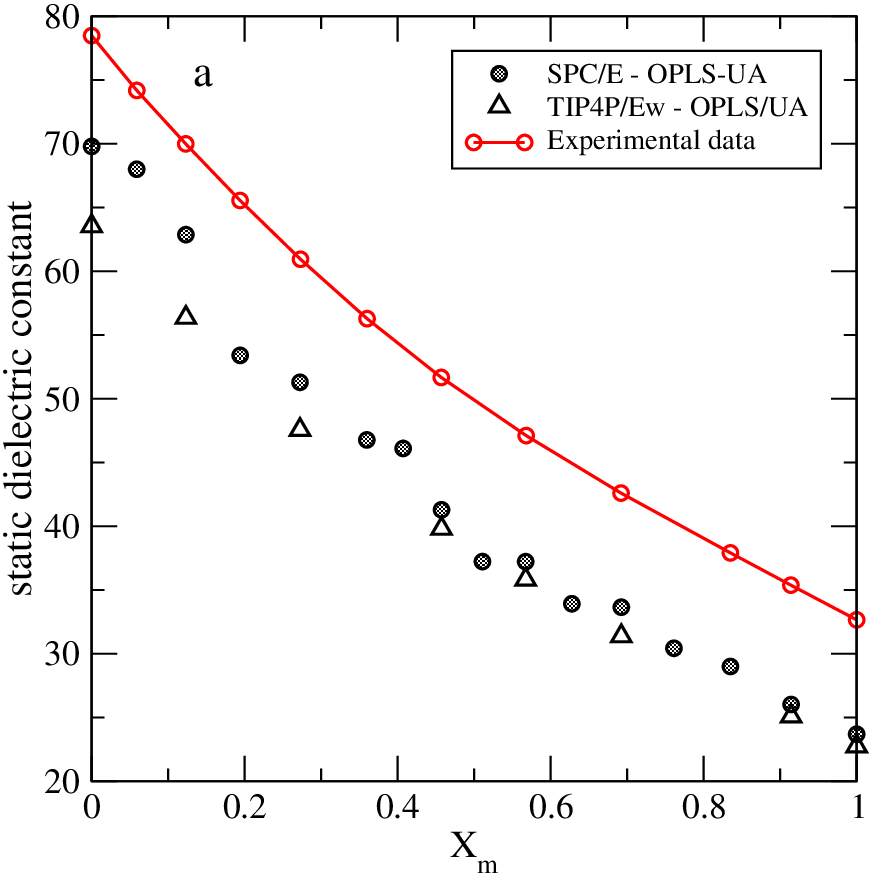}
\hspace{3mm}
\includegraphics[width=0.46\textwidth,clip]{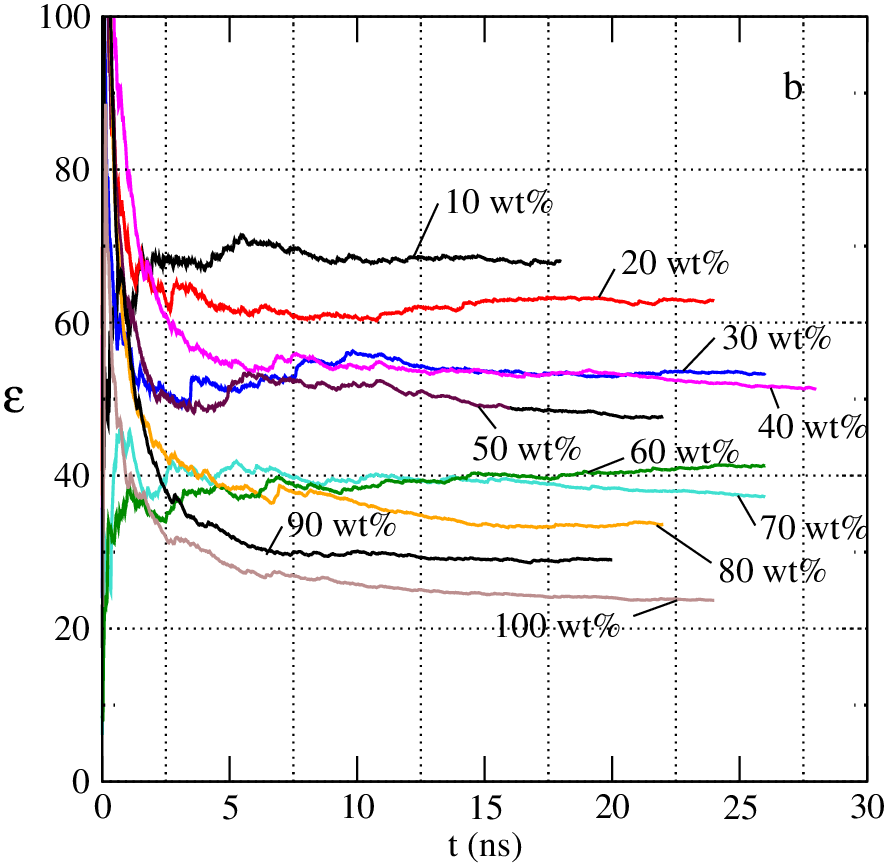}
\end{center}
\vspace{-3mm}
\caption{(Color online) Panel (a): Composition dependence of the dielectric constant of
water-methanol mixture
from our present simulations versus experimental data  \cite{albright}.
The nomenclature of lines and symbols is given in the inset.
Panel (b): Time dependence of the dielectric constant at different weight concentration.
\label{fig4}}
\protect
\end{figure}

Now  we proceed to another property that, as seen below, is not very successfully
described within the framework
of the potential models applied.
It is known that the long-range, asymptotic behavior of correlation functions
between molecules possessing dipole moments is described by the dielectric constant, $\varepsilon$.
In general terms, the calculation of the dielectric constant from
simulations is a demanding task, see e.g.,  \cite{alejandre1,max,gereben} for the discussion
of calculations of $\varepsilon$ for pure water. Several
factors \textit{can} influence the result, such as  the number of molecules, type of thermostat and
barostat, precision of the summation of long-range interactions.
Moreover, long runs are necessary to obtain reasonable estimates for this property,
since it is usually calculated from the time-average of the fluctuations of the total
dipole moment of the system  \cite{martin} as follows:
\begin{equation}
\varepsilon=1+\frac{4\pi}{3k_\textrm{B}TV}\left(\langle{\bf M}^2\rangle-\langle{\bf M}\rangle^2\right),
\end{equation}
where $k_\textrm{B}$ is the Boltzmann constant and V is the simulation cell volume.

Both of the models for the pure species, water and methanol, are perfect in predicting their
respective dielectric constants. As a result, the behavior of mixtures in the entire
composition range is not perfect either, figure~\ref{fig4}~(a). The inclination of the curve for $\varepsilon (X_\textrm{m})$
is not bad at all, but the values are too low compared to the experimental results.
We get confidence in the quality of the present results
making long simulations [figure~\ref{fig4}~(b)] with a sufficiently large number of particles in the simulation box.
A simultaneous improvement of water and methanol models is necessary to obtain better values of the
dielectric constant for the mixtures in question. How good is the parametrization
of the methanol model of  \cite{vrabec2} with respect to the dielectric constant is unknown so far.
On the other hand, quite recently the united atom model of methanol was parameterized
to reproduce the experimental value of the dielectric constant~\cite{dominguez},
at the expense of making other properties worse, e.g., the self-diffusion coefficient.
However, the performance of this modified model for water-methanol mixtures has not been evaluated yet.
We plan to improve the description of the dielectric constant by performing
simulations for more sophisticated models. Without such efforts it is difficult to attempt
a successful description of the thermodynamics and the dynamics of ionic solvation in mixed water-alcohol
solvents.

\subsection{Dynamic properties and hydrogen bonding}

\begin{figure}[!h]
\begin{center}
\includegraphics[width=0.45\textwidth,clip]{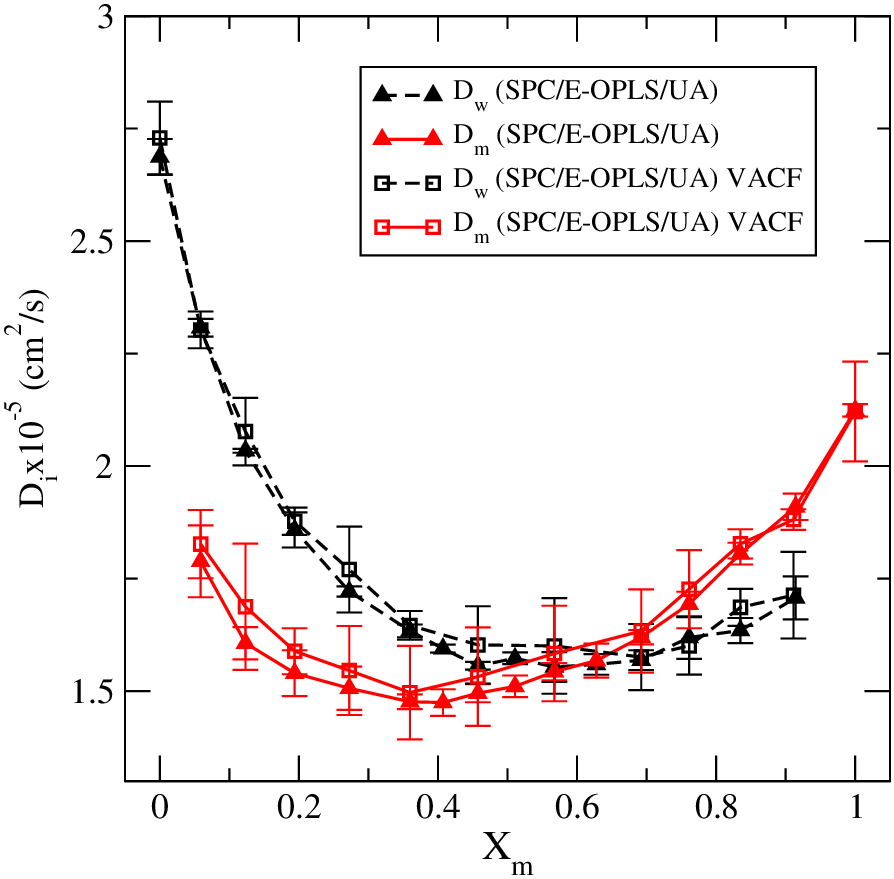}
\end{center}
\vspace{-3mm}
\caption{(Color online) A comparison of self-diffusion coefficients of the two species coming
from mean square displacements calculations and from the velocity autocorrelation functions.
Error bars are shown.
The subscripts ``m'' and ``w'' refer to methanol and water, respectively.
\label{fig5}}
\protect
\end{figure}

We proceed to the description of our results for a set of dynamic properties.
The self-diffusion coefficients for the two species of the mixture were calculated by two routes.
Namely, from the mean-square displacement (MSD) of a particle via the Einstein relation,
\begin{equation}
D_i =\frac{1}{6} \lim_{t \rightarrow \infty} \frac{\rd}{\rd t} \langle\vert {\bf r}_i(t)-{\bf r}_i(0)\vert ^2\rangle,
\end{equation}
where $i$ refers to water or methanol species. The MSD procedure to obtain $D_i$ was applied
by using particle coordinates coming from DLPOLY simulations.
On the other hand, we also obtained the self-diffusion coefficients
from the time integral of the velocity autocorrelation functions,
\begin{equation}
D_i =\frac{1}{3} \int_0 ^\infty \rd t \langle {\bf v}_i(t){\cdot}{\bf v}_i(0)\rangle.
\end{equation}
For this purpose we employed the GROMACS software for the same model and with the same technical
setup as in the DLPOLY package. This has been done due to the speed efficiency of GROMACS. For the sake
of checking mutual consistency, a set of calculations has been performed and the results are given
in figure~\ref{fig5}. We observe that the two procedures yield very similar results, margins of inaccuracy overlap
for all compositions.

\begin{figure}[!t]
\begin{center}
\includegraphics[width=0.45\textwidth,clip]{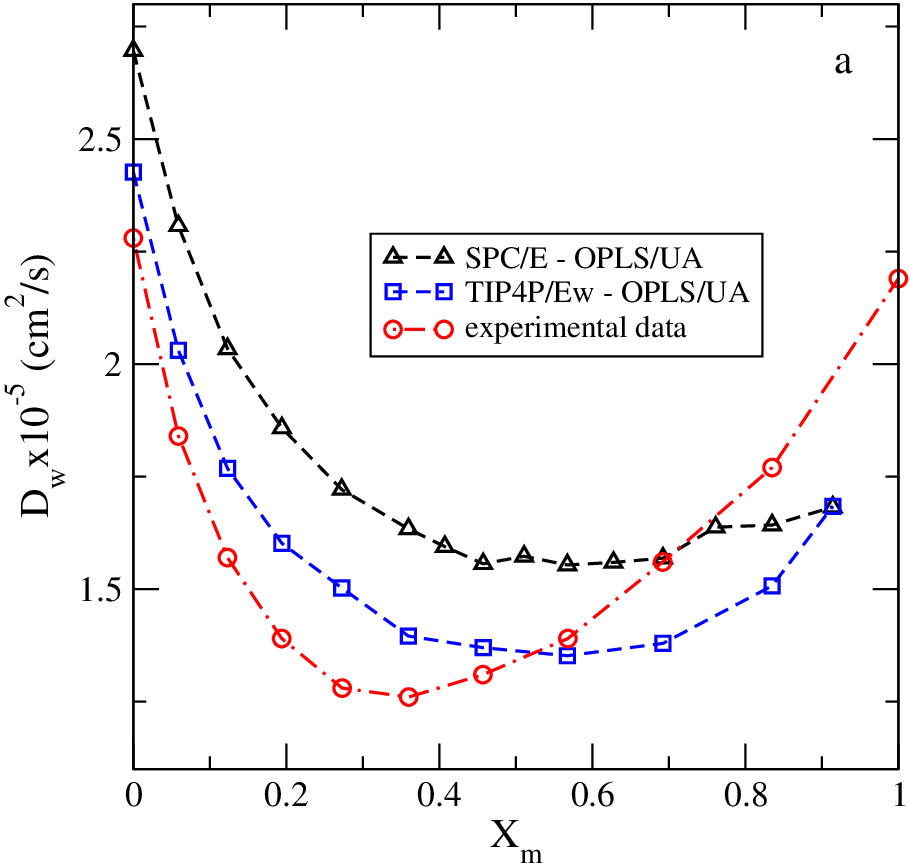}
\hspace{3mm}
\includegraphics[width=0.45\textwidth,clip]{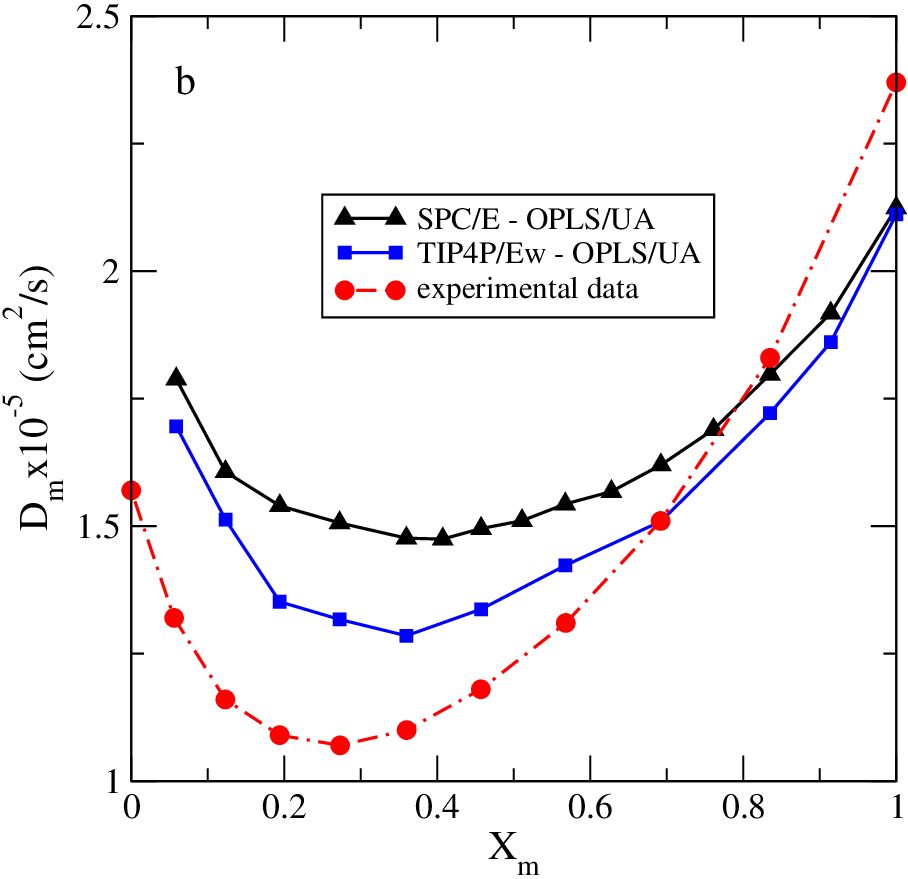}
\end{center}
\vspace{-3mm}
\caption{(Color online) Composition dependence of the self-diffusion coefficient of species
in water-methanol mixtures from the present simulations in comparison
with experimental data  \cite{derlacki}.
\label{fig6}}
\end{figure}

A set of results for self-diffusion coefficients coming from simulations by using VACF
is given in figure~\ref{fig6}. While an overall behavior of the
curves for water and methanol molecules is correct, the experimental  \cite{derlacki} and simulation data
differ in several details. On the water-rich side, the simulation data overestimate the self-diffusion
coefficients as a result of inability of the explored models for water to yield the self-diffusion
coefficient in perfect agreement with the experimental value for pure water.
The self-diffusion coefficient for pure methanol is not as bad. The OPLS/UA model slightly
underestimates the self-diffusion coefficient value. However, the minima of the curves
for each species of the mixtures are less pronounced than in the experiment. Moreover,
the crossing point where the self-diffusion coefficient of water and methanol molecules are
equal is predicted by the simulation at a high $X_\textrm{m}$ ($X_\textrm{m} \approx 0.6$), whereas
in the experiment this crossing occurs at an even higher methanol fraction, around~0.7.

It is not straightforward to establish any correlation between the behavior of mixing properties
and the self-diffusion coefficients of the species. Experimentally, the minimum values of excess
mixing volume and enthalpy, as well as of self-diffusion coefficients, occur at a slightly prevailing
water fraction, cf. figures~\ref{fig2}, \ref{fig3}  and \ref{fig6}.
As concerns the excess volume, visually it looks as if the minimum position was at $X_\textrm{m}=0.5$.
On the other hand, simulations involving the OPLS/UA methanol model
indicate that the minima for $\Delta H_\textrm{mix}$ and $D_w$ are located at a prevailing methanol fraction.
In order to explore
this discrepancy more in  detail we resort to the notion of hydrogen bonding.

The structural, mixing and dynamic properties in hydrogen bonded liquids are all
certainly locally  connected with the characteristics
of the H-bonds, and globally with the nature of the hydrogen bonded network that is formed
in a given system.
As a step towards understanding this connection, we have calculated characteristics
related to H-bonding.
We decided to use a rather popular geometric criterion applicable to water and methanol
that involves two distances and one angle,
see e.g.,  \cite{kumar,zhang1}.
Three conditions determine the existence of an H-bond, namely the distance $R_{\textrm{O}_x\textrm{O}_y}$ between
oxygen atoms belonging to two
molecules should not exceed the
threshold value $R_{\textrm{O}_x\textrm{O}_y}^\textrm{c}$; the distance $R_{\textrm{O}_y\textrm{H}_x}$ between the acceptor oxygen
atom and the hydrogen
atom connected with the donor oxygen should not exceed the threshold value $R_{\textrm{O}_y\textrm{H}_x}^\textrm{c}$. Finally, the
angle $\textrm{O}_x - \textrm{H}_x \cdots \textrm{O}_y$ should be smaller than the threshold value,
see e.g., references  \cite{padro1,guardia1,guardia2} concerning the analysis of the
structure of liquid methanol.
If the coordinates of two molecules fulfill the above conditions, they are considered to be H-bonded.
The values used in this work are the natural
ones coming from the first (intermolecular) minima of the corresponding radial distribution functions for atoms O--O and O--H
belonging to different species
at each calculated composition. There is a certain freedom concerning the choice of a threshold value for the angle.
We have chosen to use the most popular threshold value for the
angle between two vectors, $\widehat{{\bf r}_{\textrm{O}_x\textrm{O}_y}{\bf r}_{\textrm{O}_x\textrm{H}_x}}$,
equal to $30^\circ$ that yields, as we will see below, a higher fraction of H-bonded molecules in
comparison with the equally successful, more restrictive criterion for the angle between vectors,
$\widehat{{\bf r}_{\textrm{H}_x\textrm{O}_x}{\bf r}_{\textrm{H}_x\textrm{O}_y}}$,
larger than  $150^\circ$, see the angle $\theta$ in
figure~1 of reference  \cite{kumar}.

\begin{figure}[!t]
\begin{center}
\vspace{2mm}
\includegraphics[width=0.44\textwidth,clip]{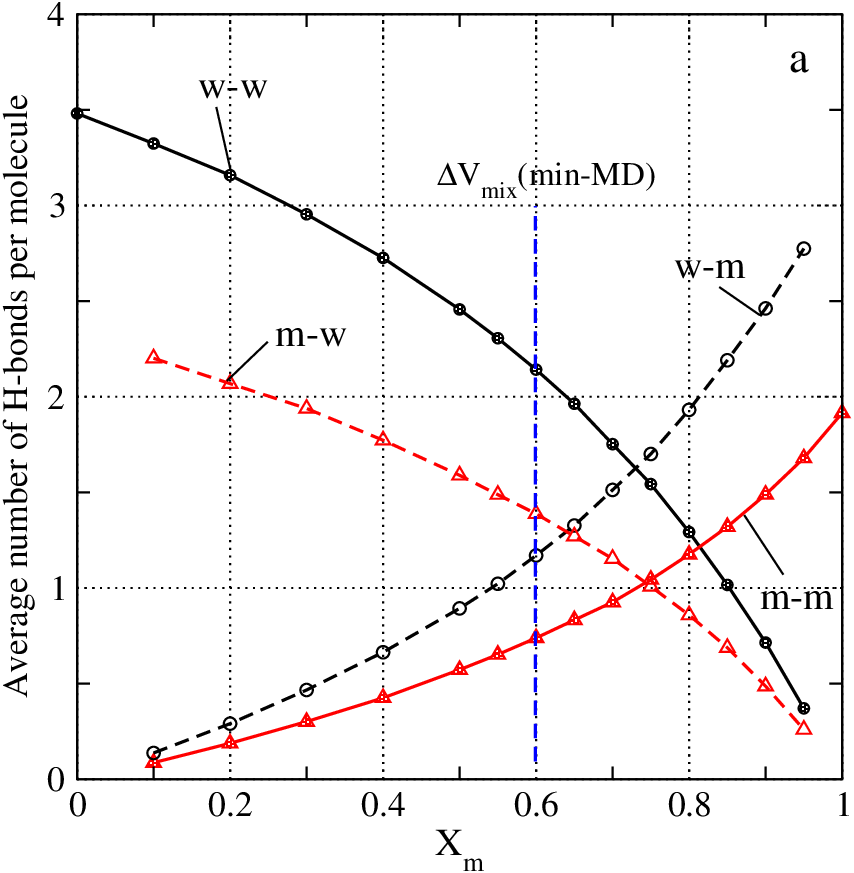}
\hspace{3mm}
\includegraphics[width=0.45\textwidth,clip]{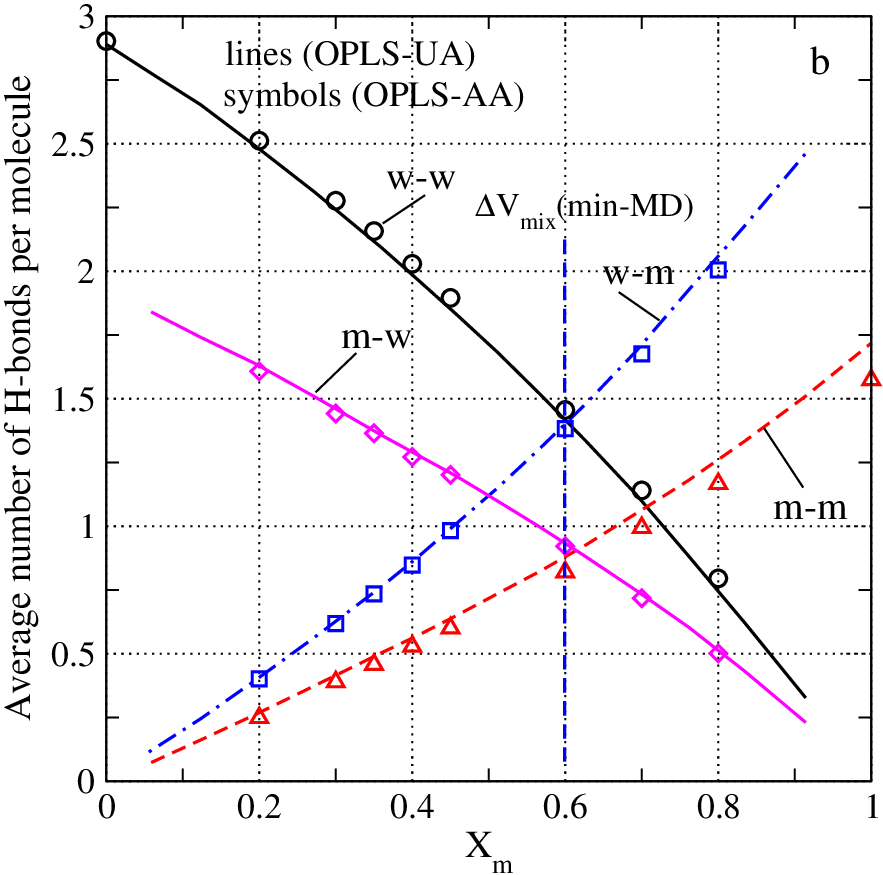}
\end{center}
\vspace{-2mm}
\caption{(Color online) Panel (a): The average number of H-bonds
per molecule of different species for SPC/E water model combined with united atom
modelling of methanol (threshold angle criterion~--- $30^\circ$, see description in the text).
Panel (b): Same quantities as in panel (a), with a different geometric criterion for H-bonds between
molecules, namely $\theta = 150^\circ$ as a threshold angle. The OPLS/AA data are given
for the sake of comparison. Normalization of the data is explained in the text.
\label{fig7}}
\protect
\end{figure}

In the two panels of figure~\ref{fig7}, the average number of hydrogen bonds per molecule is given as a function of composition for two different geometric definitions of H-bonds. Both definitions lead to similar trends whereas the actual numbers at each composition are a bit different.
The values are normalized  per the number of molecules of the first
species marked next to a given line (w~--- water; m~--- methanol).
That is, for instance, the line ``w--m'' gives the number
of hydrogen bonds, $\langle n \rangle _\textrm{HB}$, for a water molecule that forms connections with methanol ones
(normalized by the number of water molecules).
In figure~\ref{fig7}~(b), the values for the UA and AA models of methanol are also compared:
correspondence between the two types of potentials is nearly perfect in this respect.
In both panels of figure~\ref{fig7} the vertical dashed line marks an approximate composition ($X_\textrm{m} \approx 0.60$) at which
the excess mixing volume is at minimum, cf. figure~\ref{fig2} for TIP4P-OPLS-UA combination. The other
two water models appear to put the minimum position to about 0.5.
The behavior of $\Delta V_\textrm{mix}$ is well
correlated with the average number of H-bonds. The line ``w--w'' decays in the entire range of composition
reflecting the shrinking of the system due to breaking of H-bonds, whereas the ``w--m'' line shows an increasing average number of
bonds and leading to ``expansion'' of the system due to an increasing number of bonds.
The ``m--m'' and ``m--w'' lines exhibit similar trends. Consequently, in two regions where one of the two
average numbers of bonds prevail, the
excess mixing volume decreases or increases. In the intermediate interval of compositions, where
the average numbers of bonds are balanced, $\Delta V_\textrm{mix}$ exhibits its minimum. These trends
are independent  of the definition of the H-bond. If $\Delta V_\textrm{mix}$ is in the interval of its minimum
value, the self-diffusion coefficients are also approximately around their respective minima.
This picture comes from computer simulations of both of the two water models, combined with OPLS/UA methanol;
we show only the case of SPC/E to avoid an unnecessary repetition. It is worth mentioning that the
excess mixing enthalpy is not straightforward to incorporate into the interpretation of the above data.
The reason is that $\Delta H_\textrm{mix}$ has ``geometric'' and energetic terms. Therefore, combining them to
interpret the observed trends using a purely geometric definition of H-bonds does not seem to be appropriate.

One interesting conjecture coming from the results in figure~\ref{fig7} is
that as a general tendency, both water and methanol molecules prefer to coordinate water
molecules via H-bonding, similarly to what was found for the mixtures with AA methanol
molecules  \cite{pusztai1}.
This can be discerned from the non-linear dependence of the number of H-bonds per molecule
on the methanol concentration.
In the non-preferential case, the number of both water and methanol H-bonded neighbors would change
linearly with concentration. The deviation from linearity is positive for
the ``w--w'' and ``m--w'' curves, i.e.,
there are more water molecules hydrogen bonded to both water and methanol molecules
than it would be proportional to the composition. By contrast, the deviation is negative
for the ``w--m'' and ``m--m'' curves, i.e.,  the number of H-bonded methanol molecules to both water
and methanol molecules  is smaller than it would be proportional
to the methanol concentration. Interestingly, the effect is more pronounced for the
less strict definition of H-bonds, as exemplified by figure~\ref{fig7}~(a).

\begin{figure}[!t]
\begin{center}
\includegraphics[width=0.45\textwidth,clip]{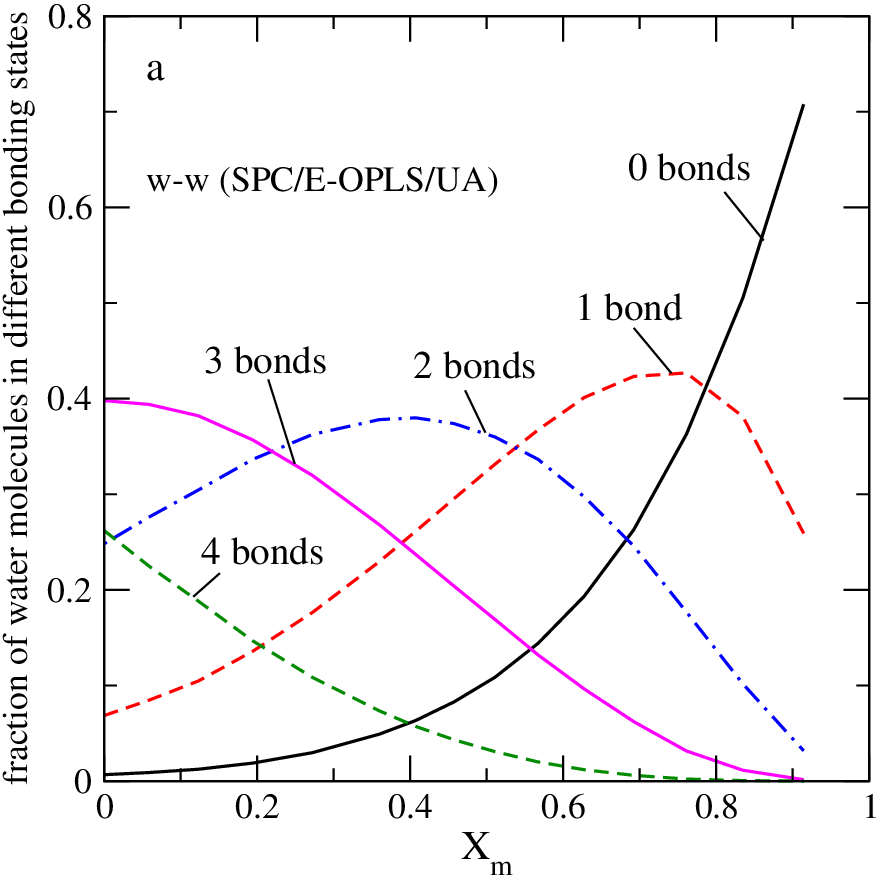}
\hspace{3mm}
\includegraphics[width=0.45\textwidth,clip]{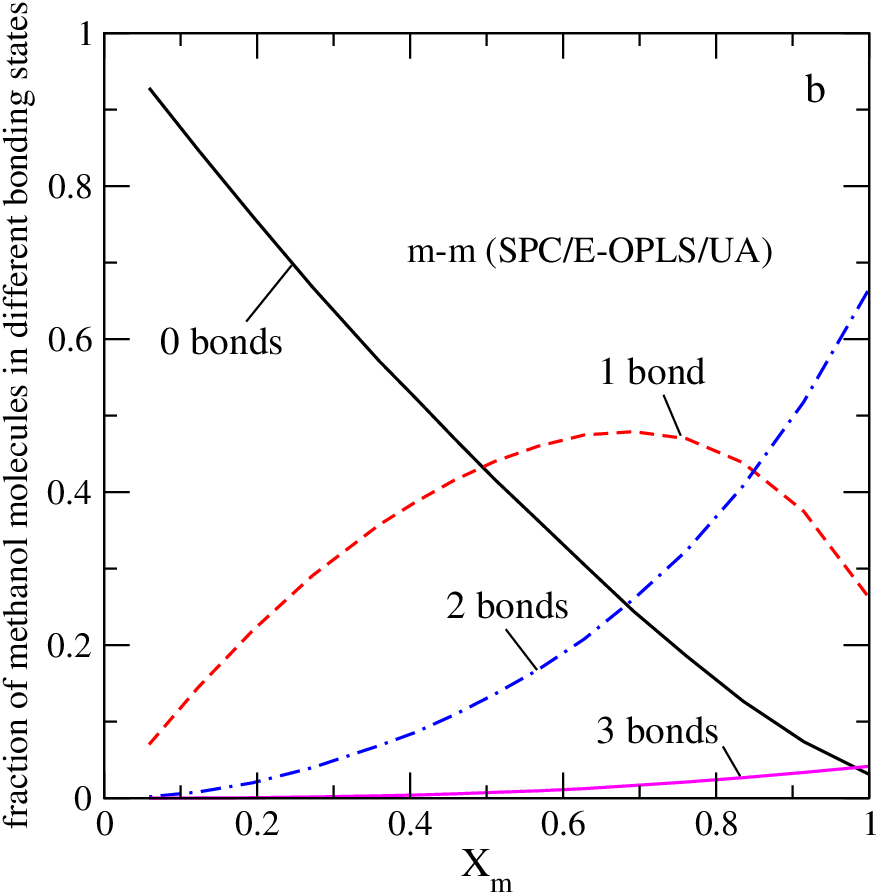} \\[2pt]
\includegraphics[width=0.45\textwidth,clip]{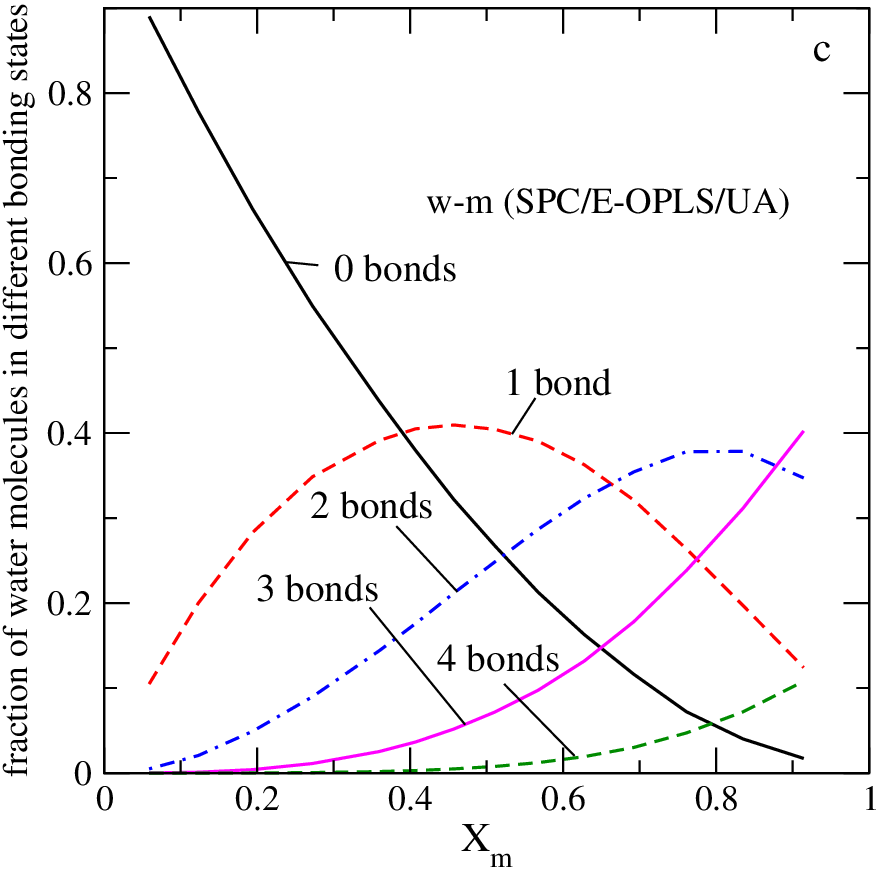}
\hspace{3mm}
\includegraphics[width=0.45\textwidth,clip]{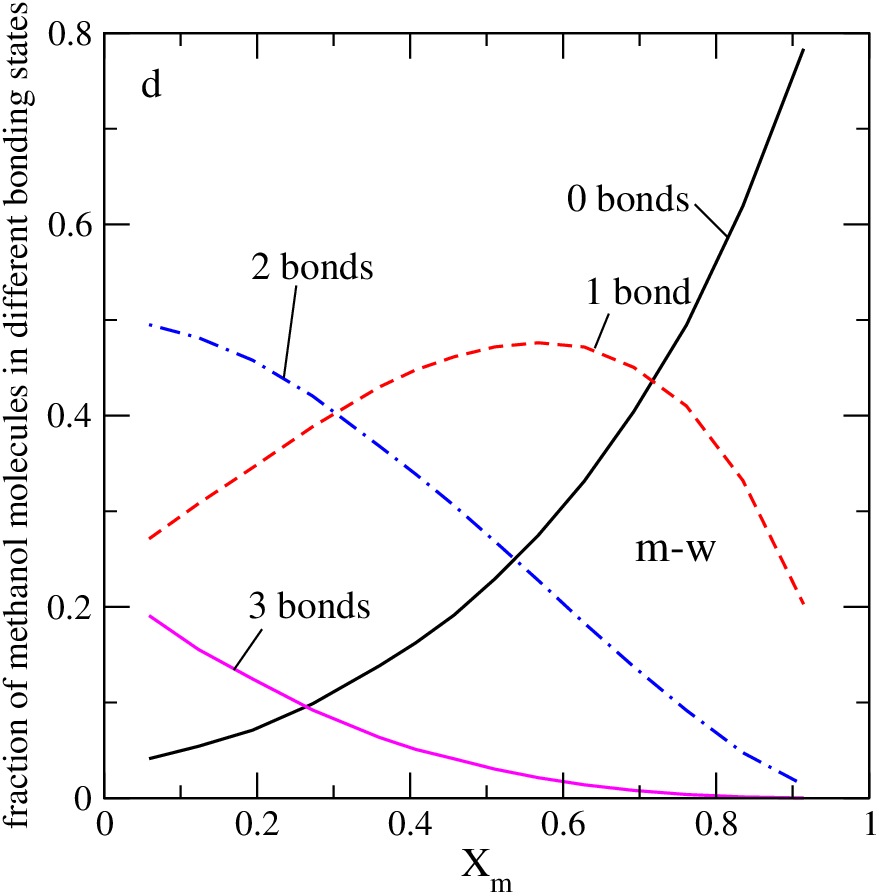}
\end{center}
\vspace{-3mm}
\caption{(Color online) A comparison of the fractions of differently bonded
molecules of water and methanol for united atom
modelling of methanol, panels (a), (b), (c) and (d), respectively. The criterion $\theta = 150^\circ$ is used.\label{fig8}}
\protect
\end{figure}

By a close inspection of the fractions of methanol and water molecules in different bonding states,
as a function of composition (see figure~\ref{fig8}), a very similar conclusion may be drawn.
These graphs provide the ratio of water or methanol molecules that have a specific
number of H-bonded water
or methanol neighbors at a given composition. Water can form a maximum of four hydrogen bonds,
whereas methanol can form three. If we compare water-water [``w--w'' in figure~\ref{fig8}~(a)] and
water-methanol [``w--m'' in figure~\ref{fig8}~(c)]  curves, the following observations can be made.
As concerns panel (a) of figure~\ref{fig8} that describes the bonding solely within the water subsystem of the
mixture, the following trends are worth mentioning. Namely,
when the methanol concentration increases (starting from pure water up to pure methanol), the fractions
of water molecules with four and three ``w--w'' bonds monotonously decrease whereas the fraction of
molecules that do not participate in bonding substantially grows. Only the fractions of doubly
and singly bonded water molecules exhibit maxima at a respective composition, $X_\textrm{m}$. In other words,
clusters (and branched structures) of exclusively water molecules transform into chains, and thus dimers as
well as ``free'' molecules become abundant. This is a crude description of how a water subsystem
of a mixture shrinks. Actually, the minimum value of the excess mixing volume is located in between the
two maxima observed for doubly and singly bonded waters.

On the other hand, see figure~\ref{fig8}~(c), the fractions describing the bonding of water molecules solely with
methanol species exhibit the following trends. The average number of waters initially free of
``w--m'' bonds rapidly decreases at the expense of the formation of bonded structures. Only the fractions
of water molecules with a single bond and double bonds with methanols pass through a maximum
value, other fractions with higher number of bonds grow monotonously. Thus, the expansion of the
water subsystem occurs via substitution of water molecules by methanols in a shell of neighbors of
a given water molecule. In close similarity to what we have seen in panel (a), the minimum of
the excess mixing volume occurs at $X_\textrm{m}$ located in between two maxima of the curves describing singly
and doubly bonded molecules. Similar interpretation can be developed to describe the changes of the
bonding within the methanol subsystem and for the changes of the fraction of ``m--w'' bonds, panels (b) and
(d) of figure~\ref{fig8}.

\begin{figure}[!h]
\begin{center}
\includegraphics[width=0.45\textwidth,clip]{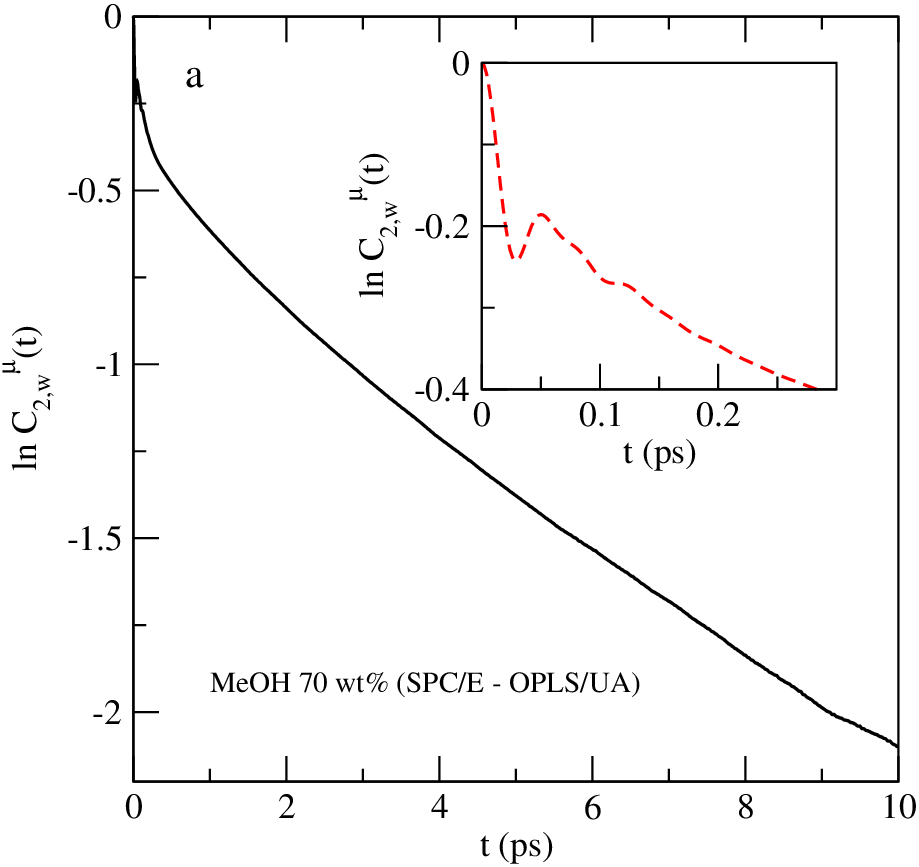}
\hspace{3mm}
\includegraphics[width=0.45\textwidth,clip]{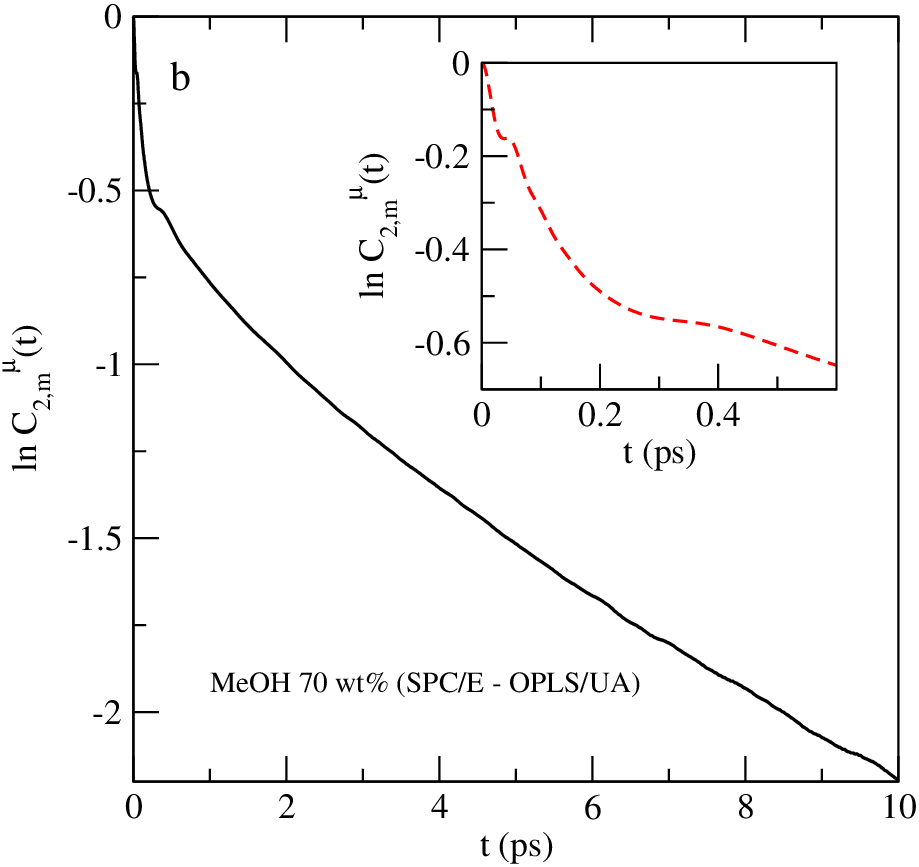}
\end{center}
\vspace{-3mm}
\caption{(Color online) An example of the behavior of the orientational auto-correlation functions of
the two species in the mixture in the dipole frame. Left-hand panel: water; right-hand panel: methanol.\label{fig9}}
\protect
\end{figure}

Our final focus in the present study is in the relaxation properties of molecules in mixtures.
Those certainly are also influenced by H-bonding discussed above.
In close similarity to the previous studies  \cite{wensink,spoel}, we would like to evaluate
the reorientational correlation functions determined from
\begin{equation}
C^{\alpha}_l(t) = \langle P_l[{\bf e}^{\alpha}(t){\cdot}{\bf e}^{\alpha}(0)]\rangle,
\end{equation}
where $P_l$ is the $l$th order Legendre polynomial and ${\bf e}^{\alpha}$ is the unit vector which
points along the $\alpha$ axis in the molecular reference frame.
The average is performed over all molecules belonging to a particular species.
A typical plot for a mixture of a given composition describing the dependence of the
autocorrelation function on time is given in figure~\ref{fig9}.

In the present calculations we used
the axis parallel to the molecular dipole moment $\mu$ which is related to dielectric relaxation.
Moreover, we calculated $C^{\mu}_l$ for $l=2$ for the two species of the mixture.
In addition, we performed calculations by using the O--H vector (H is the hydrogen atom belonging to the
hydroxyl group of the molecule) as $\alpha$ axis to obtain the corresponding relaxation time.
The single-molecule reorientational times follow from integration of the autocorrelation
function defined above:
\begin{equation}
\tau^{\alpha}_l = \int_0 ^\infty \rd t C^{\alpha}_l(t).
\end{equation}
To perform the integration we split the time interval into two pieces,
actually it is a commonly accepted procedure  \cite{laaksonen,spoel}.
Over a short-time interval, the integration is performed numerically; the upper limit
of this interval is chosen between 4 and 5 ns approximately.
For a long-time interval, the logarithm of the autocorrelation function is plotted:
it behaves as a straight line permitting to analytically evaluate the relevant contribution
 \cite{spoel}.

\begin{figure}[!t]
\begin{center}
\includegraphics[width=0.45\textwidth,clip]{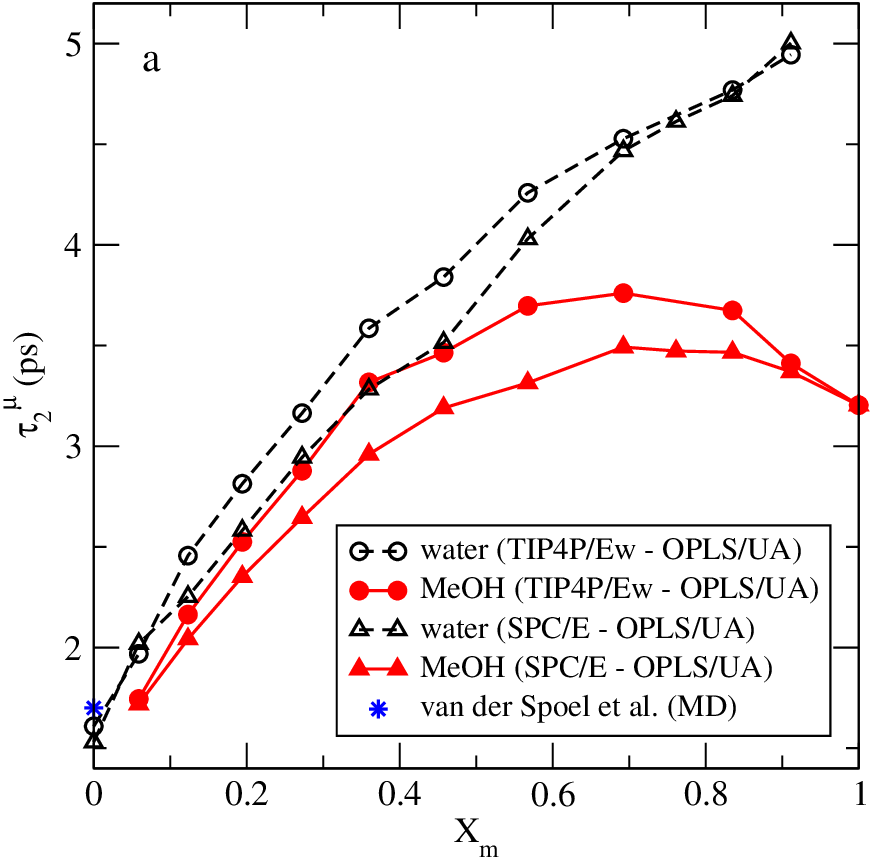}
\hspace{3mm}
\includegraphics[width=0.45\textwidth,clip]{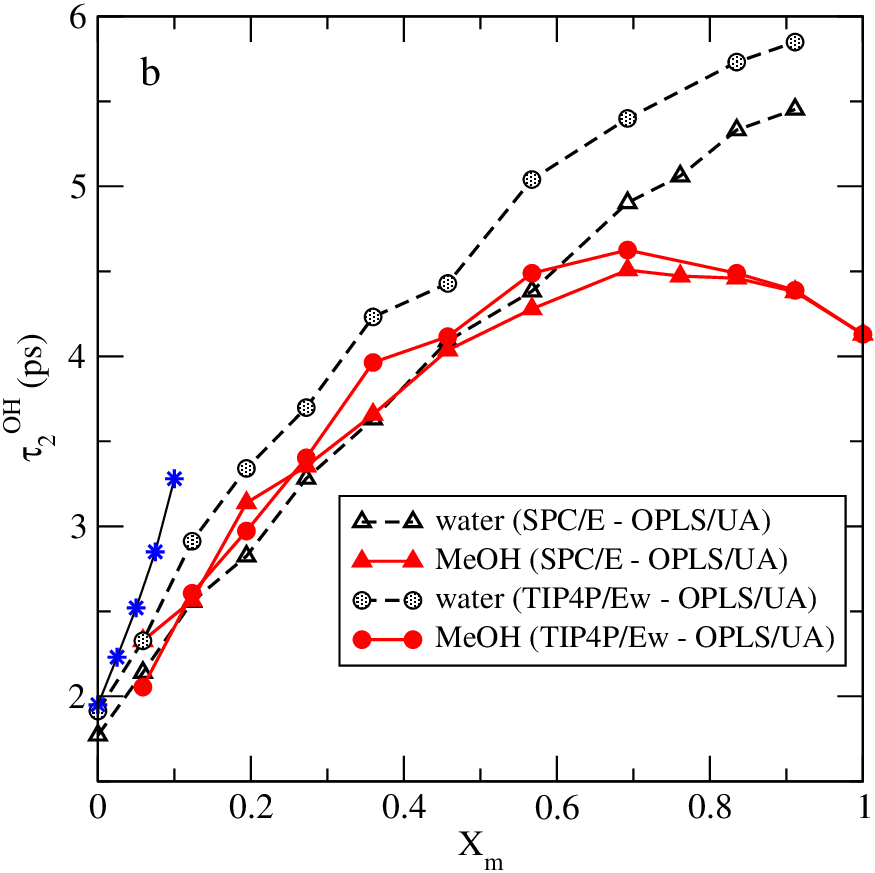}
\end{center}
\vspace{-3mm}
\caption{(Color online) The behavior of rotational relaxation times in the
dipole frame [panel~(a)] and in the OH frame [panel~(b)], as a function of the composition of the
mixture. The experimental data given as a blue line and stars in panel~(b) are from \cite{ludwig}.\label{fig10}}
\protect
\end{figure}

Single molecule reorientational times, as calculated for both kinds of the axes
mentioned above, are reported in figure~\ref{fig10}, as a function of composition.
Two trends are qualitatively similar in the behavior of $\tau^{\mu}_{2,i}$ and $\tau^\textrm{OH}_{2,i}$ on
composition. Namely, both relaxation times characterizing the water species grow with an increasing $X_\textrm{m}$,
showing a slower dynamics of water molecules in methanol-rich mixtures compared to mixtures with
low amount of methanol. On the other hand, the two relaxation times characterizing methanol species
in a mixture have maxima, at $X_\textrm{m} \approx 0.70$.
The slow dynamics, i.e., high relaxation time observed
in pure methanol turns even slower at $X_\textrm{m} \approx 0.70$; passing this maximum, the relaxation times monotonously
drop, showing a faster dynamics of methanol species for mixtures in which the water component is
predominant. Such trends are observed for both water models in question when combined with the methanol OPLS/UA
model. Very similar trends and a qualitatively similar behavior has been discussed by \cite{wensink}
for $\tau^{\mu}_{2,i}$ on methanol fraction within the framework of TIP4P-OPLS/AA model. Wensink et al. values,
however, are substantially lower in comparison with our results with united atom modelling.
It is difficult to establish how accurate all these results are due to a lack of well
established  experimental data. We included some
points obtained by Ludwig  \cite{ludwig} in figure~\ref{fig10}~(b). At present it seems impossible
to perform a comprehensive validation of different models versus such kind of experimental data.

\section{Summary}

To summarize this study, we have used
molecular dynamics simulations in the NPT ensemble to study some thermodynamic properties,
as well  as the dynamic and
dielectric properties of water-methanol mixtures at room temperature and at ambient pressure
in the entire range of their composition. The SPC/E and TIP4P/Ew models combined with the
OPLS/UA potential model for methanol are used. This is done as a first step of systematic studies
of mixtures of water and a set of alcohols using nonpolarizable models.
Our principal findings concern the behavior of density, excess mixing volume and excess
mixing enthalpy on composition. Also, we have evaluated the self-diffusion coefficients
of both species, reorientational relaxation times and trends of behavior of hydrogen bonding
between molecules. Whenever it seemed possible, we looked for a unified interpretation of different
properties and performed comparisons with available experimental results.
It would be interesting to establish a relation between the structural properties and
hydrogen bonding on the one hand and the dielectric constant on the other hand,
possibly in the spirit of previous studies of diffusion and power spectra,
see e.g.,  \cite{guardia3}.
We would like to note that in our opinion an ample set of properties of the model mixtures
in question is required to establish if the modelling is consistent (with experimental data)
and successful. A combination of experimental data, computer simulation results and
reverse Monte Carlo modelling along the lines developed in  \cite{pusztai2,pusztai3} may be profitable
in this respect.
Possible modifications of parameters and combination rules nevertheless should be explored more in detail.
At present, it seems that each specific parametrization leads to a successful description of a particular
property, such state of art being unsatisfactory. A more sophisticated modelling of alcohols,
e.g., OPLS/AA, and the use of possibly more refined water models is necessary
to extend the present study and at least make stronger conclusions about  various
thermodynamic properties of these mixtures.  Unfortunately,  it seems that some of them are
intrinsically impossible to describe very well within the framework of non-polarizable models.

\section*{Acknowledgements}

We acknowledge support of CONACyT of M\'exico (grant Nu.~214477) and the National Research,
Development and Innovation Office of Hungary (grant no.~T\'ET\_12\_MX-1-2013-0003)
for funding a bilateral
collaboration between the Mexican and Hungarian partners, and of DGAPA de la UNAM
under project IN-205915. E.G. was supported by CONACyT of Mexico under a Ph.D. scholarship.
E.G. and O.P. are grateful  to Dr.~T.~Patsahan for helpful discussions and valuable comments.
O.P.~is grateful  to D.~Vazquez and M.~Aguilar for technical support of this work.

\ukrainianpart

\title{Композиційна залежність термодинамічних, динамічних та діелектричних властивостей модельних сумішей вода-метанол. Результати симуляцій методом молекулярної динаміки}

\author{E. Галіція-Андрес\refaddr{label1}, Г. Домінгес\refaddr{label2},
Л. Пустаї\refaddr{label3}, О. Пізіо\refaddr{label1}}
\addresses{
\addr{label1}Інститут хімії, Національний автономний університет м. Мехіко,
Мехіко, Мексика
\addr{label2} Інститут матеріалознавства, Національний автономний університет м. Мехіко,
Мехіко, Мексика
\addr{label3}  Фізичний дослідницький центр Вігнера, Угорська академія наук,
Будапешт, Угорщина}

\makeukrtitle

\begin{abstract}
Нами досліджено термодинамічні та динамічні властивості, а також діелектричну сталу модельних сумішей вода-метанол  у всьому діапазоні концентрацій з використанням симуляцій методом молекулярної динаміки при постійному тиску і в довкільних умовах. Використано моделі води SPC/E і  TIP4P/Ew
у поєднанні  з OPLS об'єднаним атомним моделюванням для метанолу. Зміни середнього числа водневих зв'язків між частинками різних сортів та фракцій по-різному зв'язаних молекул поставлено у відповідність з поведінкою   об'єму надлишкового змішування та ентальпії, коефіцієнтів самодифузії та
часів ротаційної релаксації. Детально проаналізувавши отримані в роботі результати, робимо висновок, що можна досягнути певного вдосконалення опису великої кількості властивостей  сумішей вода-метанол за умови використання більш складної, ретельно параметризованої, наприклад, повністю атомної моделі. До того ж, існує потреба у дослідженні параметризації силового поля метанолу з одночасним використанням різних комбінаційних правил для взаємодій метанол-вода.

\keywords моделі води, моделі метанолу, термодинамічні властивості, трансляційна та орієнтаційна дифузія, діелектрична стала, молекулярна динаміка

\end{abstract}


\begin{thebibliography}{99}

\bibitem{pusztai1}  Galicia Andr\'es E.,  Pusztai L.,  Temleitner L.,  Pizio O., J. Mol. Liq., 2015, \textbf{209}, 586;
\doi{10.1016/j.molliq.2015.06.045}.

\bibitem{spce}  Berendsen H.J.C.,  Grigera J.R.,  Straatsma T.P.,
J. Phys. Chem., 1987, \textbf{91}, 6269; \doi{10.1021/j100308a038}.

\bibitem{tip4p-ew}  Horn H.W.,  Swope W.C.,  Pitera J.W.,  Madura J.D.,  Dick T.J.,  Hura~G.L.,  Head-Gordon~T., J. Chem. Phys., 2004, \textbf{120}, 9665; \doi{10.1063/1.1683075}.

\bibitem{opls-aa}  Jorgensen W.L.,  Maxwell D.,  Tirado-Rives S., J. Am. Chem. Soc.,
1996, \textbf{118}, 11225; \doi{10.1021/ja9621760}.

\bibitem{jorgensen1} Jorgensen W.L.,  Madura J.D., J. Am. Chem. Soc., 1983, \textbf{105}, 1407; \doi{10.1021/ja00344a001}.

\bibitem{okazaki1} Okazaki S.,  Nakanishi K.,  Touhara H., J. Chem. Phys., 1983, \textbf{78}, 454; \doi{10.1063/1.444525}.

\bibitem{okazaki2} Okazaki S.,  Touhara H.,  Nakanishi K., J. Chem. Phys., 1983, \textbf{81}, 890; \doi{10.1063/1.447726}.

\bibitem{ferrario}  Ferrario M.,  Haughney M.,  McDonald I.R.,  Klein M.L.,
J. Chem. Phys., 1990, \textbf{93}, 5156; \doi{10.1063/1.458652}.

\bibitem{tanaka}  Tanaka H.,  Gubbins K.E., J. Chem. Phys., 1992, \textbf{97}, 2626; \doi{10.1063/1.463051}.

\bibitem{laaksonen}  Laaksonen A.,  Kusalik P.G.,  Svishchev I.M., J. Phys. Chem., 1997, \textbf{101}, 5910; \doi{10.1021/jp970673c}.

\bibitem{gabor1} Palinkas G., Hawlicka E.,  Heinzinger K., Chem. Phys., 1991, \textbf{158}, 65; \doi{10.1016/0301-0104(91)87055-Z}.

\bibitem{gabor2} Palinkas G., Bako I.,  Heinzinger K.,  Bopp P., Mol. Phys., 1991, \textbf{73}, 897; \doi{10.1080/00268979100101641}.

\bibitem{gabor3} Palinkas G., Bako I., Z. Naturforsch. A, 1991, \textbf{46}, 95; \doi{10.1515/zna-1991-1-215}.

\bibitem{tip4p} Jorgensen W.L., Chandrasekhar J., Madura J.D., Impey R.W., Klein M.L.,
J. Chem. Phys., 1983, \textbf{79}, 926; \\ \doi{10.1063/1.445869}.

\bibitem{wensink}  Wensink E.J.W.,  Hoffmann A.C.,  van Maaren P.J.,  van der Spoel D.,
J. Chem. Phys., 2003, \textbf{119}, 7308; \\ \doi{10.1063/1.1607918}.

\bibitem{vrabec1}  Guevara-Carrion G.,  Vrabec J.,  Hasse H., J. Chem. Phys.,
2011, \textbf{134}, 074508; \doi{10.1063/1.3515262}.

\bibitem{vega}  Abascal J.L.,  Vega C., J. Chem. Phys., 2005, \textbf{123}, 234505; \doi{10.1063/1.2121687}.

\bibitem{vrabec2}  Schnabel T.,  Srivastava A.,  Vrabec J.,  Hasse H., J. Phys. Chem. B, 2007, \textbf{111}, 9871; \doi{10.1021/jp0720338}.

\bibitem{nezbeda1} Gonz\'alez-Salgado D., Nezbeda I., Fluid Phase Equilib., 2006, \textbf{240}, 161; \doi{10.1016/j.fluid.2005.12.007}.

\bibitem{nezbeda2}  Mou\v{c}ka F., Nezbeda I., J. Mol. Liq., 2011, \textbf{159}, 47; \doi{10.1016/j.molliq.2010.05.005}.

\bibitem{alejandre1}   Alejandre J.,  Chapela G.A.,  Saint-Martin H.,  Mendoza N.,
Phys. Chem. Chem. Phys., 2011, \textbf{13}, 19728; \\ \doi{10.1039/c1cp20858f}.

\bibitem{alejandre2}   Fuentes-Azcatl R.,  Alejandre J., J. Phys. Chem. B, 2014, \textbf{118}, 1263; \doi{10.1021/jp410865y}.

\bibitem{dominguez} Salas F.J.,  M\'{e}ndez-Maldonado G.A.,  N\'{u}\~{n}ez-Rojas E.,
 Aquilar-Pineda G.E.,  Dom\'{i}nguez H.,  Alejandre J.,
 J. Chem. Theory Comput., 2015, \textbf{11}, 683; \doi{10.1021/ct500853q}.

\bibitem{jorgensen2} Jorgensen W.L., J. Phys. Chem., 1986, \textbf{90}, 1276; \doi{10.1021/j100398a015}.

\bibitem{vanleeuwen}  Van Leeuwen M.,  Smit B., J. Phys. Chem., 1995, \textbf{99}, 1831; \doi{10.1021/j100007a006}.

\bibitem{monica2}  Van Leeuwen M.E., Mol. Phys., 1996, \textbf{87}, 87; \doi{10.1080/00268979600100031}.

\bibitem{veldhuizen}  Veldhuizen R.,  de Leeuw S.W., J. Chem. Phys., 1996, \textbf{105}, 2828; \doi{10.1063/1.472145}.

\bibitem{forest}   Forester T.R.,  Smith W., DL-POLY Package of Molecular Simulation, CCLRC, Daresbury Laboratory, Daresbury, Warrington, England, 1996.

\bibitem{tables} Critical Tables of Numerical Data, Phys. Chem and Techn.,
Washbrun~E.W. (Ed.), Knovel, NY, 2003.

\bibitem{mcglashan}  McGlashan M.L.,  Williamson A.G., J. Chem. Eng. Data, 1976, \textbf{21}, 196; \doi{10.1021/je60069a019}.

\bibitem{lama}  Lama R.F.,  Lu B.C.-Y., J. Chem. Eng. Data, 1965, \textbf{10}, 216; \doi{10.1021/je60026a003}.

\bibitem{max}  Rami Reddy M.,  Berkowitz M., Chem. Phys. Lett., 1989, \textbf{155}, 173; \doi{10.1016/0009-2614(89)85344-8}.

\bibitem{gereben}  Gereben O., Pusztai L., Chem. Phys. Lett., 2011, \textbf{507}, 80; \doi{10.1016/j.cplett.2011.02.064}.

\bibitem{martin}  Neumann M., Mol. Phys., 1983, \textbf{50}, 841; \doi{10.1080/00268978300102721}.

\bibitem{albright}  Albright P.S.,  Gosting L.J., J. Am. Chem. Soc., 1946, \textbf{68}, 1061; \doi{10.1021/ja01210a043}.

\bibitem{derlacki}  Derlacki Z.J.,  Easteal A.J.,  Edge A.V.J.,  Wooolf~L.A.,  Roksandic~Z., J. Phys. Chem., 1985, \textbf{89}, 5318; \\ \doi{10.1021/j100270a039}.

\bibitem{kumar}  Kumar R.,  Schmidt J.R.,  Skinner J.L.,
J. Chem. Phys., 2007, \textbf{126}, 204107; \doi{10.1063/1.2742385}.

\bibitem{zhang1}  Zhang N.,  Li W.,  Chen C.,  Zuo J.,  Weng L.,
Mol. Phys., 2013, \textbf{111}, 939; \doi{10.1080/00268976.2012.760050}.

\bibitem{padro1}  Padr\'o J.A.,  Saiz L.,  Gu\`ardia E.,
J. Mol. Struct., 1997, \textbf{416},  243; \doi{10.1016/S0022-2860(97)00038-0}.

\bibitem{guardia1}  Gu\`ardia E.,  Mart\'i J.,  Padr\'o J.A.,  Saiz L.,  Vanderkooi A.V.,
  J. Mol. Liq., 2002, \textbf{96}--\textbf{97}, 3; \\ \doi{10.1016/S0167-7322(01)00342-7}.

\bibitem{guardia2}  Gu\`ardia E.,  Mart\'i J.,  Garc\'ia-Tarr\'es L.,  Laria D.,
J. Mol. Liq., 2005,  \textbf{117}, 63; \doi{10.1016/j.molliq.2004.08.004}.

\bibitem{spoel} Van der Spoel D.,  van Maaren P.J.,  Berendsen H.J.C., J. Chem. Phys., 1998, \textbf{108}, 10220; \doi{10.1063/1.476482}.

\bibitem{ludwig}  Ludwig R., Chem. Phys., 1995, \textbf{195}, 329; \doi{10.1016/0301-0104(95)00050-X}.

\bibitem{guardia3}  Mart\'i J.,  Padr\'o J.A.,  Gu\`ardia E., J. Mol. Liq., 1995,  \textbf{64}, 1; \doi{10.1016/0167-7322(95)92817-U}.

\bibitem{pusztai2} Pusztai L.,  Pizio O.,  Sokolowski S., J. Chem. Phys., 2008, \textbf{129}, 184103; \doi{10.1063/1.2976578}.

\bibitem{pusztai3} Pusztai L.,  Harsanyi I.,  Dominguez H.,  Pizio O., Chem. Phys. Lett., 2008,  \textbf{457}, 96; \doi{10.1016/j.cplett.2008.03.091}.

\end{thebibliography}
\end{document}